\documentclass[preprint,12pt]{elsarticle}

\usepackage{amsmath, amssymb}
\usepackage{caption, subcaption}
\usepackage{multirow}

\journal{Physics of the Dark Universe}

\begin{document}

\begin{frontmatter}



\title{Energy Scale of Lorentz Violation in Rainbow Gravity}


\author[a]{Nils A. Nilsson}
\ead{albin.nilsson@ncbj.gov.pl}
\author[a,b,c]{Mariusz P. D\c{a}browski}
\ead{Mariusz.Dabrowski@ncbj.gov.pl}

\address[a]{National Centre for Nuclear Research, Ho\.za 69, 00-681, Warsaw, Poland}
\address[b]{Institute of Physics, University of Szczecin, Wielkopolska 15, 70-451 Szczecin, Poland}
\address[c]{Copernicus Center for Interdisciplinary Studies, S{\l}awkowska 17, 31-016 Krak{\'o}w, Poland}

\begin{abstract}
We modify the standard relativistic dispersion relation in a way which breaks Lorentz symmetry - the effect is predicted in a high-energy regime of some modern theories of quantum gravity. We show that it is possible to realise this scenario within the framework of Rainbow Gravity which introduces two new energy-dependent functions $f_1(E)$ and $f_2(E)$ into the dispersion relation. Additionally, we assume that the gravitational constant $G$ and the cosmological constant $\Lambda$ also depend on energy $E$ and introduce the scaling function $h(E)$ in order to express this dependence. For cosmological applications we specify the functions $f_1$ and $f_2$ in order to fit massless particles which allows us to derive modified cosmological equations. Finally, by using Hubble+SNIa+BAO(BOSS+Lyman $\alpha$)+CMB data, we constrain the energy scale $E_{LV}$ to be at least of the order of $10^{16}$ GeV at 1$\sigma$ which is the GUT scale or even higher  $10^{17}$ GeV at 3$\sigma$. Our claim is that this energy can be interpreted as the decoupling scale of massless particles from spacetime Lorentz violating effects. 
\end{abstract}

\begin{keyword}
\sep Lorentz Violation
\sep Rainbow Gravity


\end{keyword}

\end{frontmatter}


\section{Introduction}

It is expected that any theory which aspires to bridge quantum theory and gravity will need to include the Planck length $\ell_P = \sqrt{\hbar G/c^3}$, where $\hbar$ is the reduced Planck constant, $G$ is Newton's gravitational constant, and $c$ is the speed of light. This characteristic length is derived by dimensional considerations of the constants which should appear in a regime where quantum theory, relativity, and gravity all are significant. 
It is expected that the Planck length is the minimum length which one can measure in a meaningful way. Associated with the Planck length is the Planck energy $E_{Pl} = \sqrt{\hbar c^5/G}$, which is simply the energy of a photon with de Broglie wavelength $\ell_P$.
The concept of a minimum length lies at the heart of approaches to quantum gravity such as string theory and loop quantum gravity, and has inspired a lot of theoretical work~\cite{AmelinoCamelia:1996pj, AmelinoCamelia:2000ge, AmelinoCamelia:2000mn, deBrito:2016fqe, Abbasvandi:2016oyw, Tawfik:2012hz}. The idea of spacetime foam was put forth in~\cite{Wheeler:1957mu} and has inspired research since then. According to this idea, quantum effects make spacetime nontrivial at small scales (the Planck scale), where particle-antiparticle pairs are continuously created and annihilated, curving spacetime at extremely small length- and time scales. This ''chaotic'' picture inspired the term ''spacetime foam'', or ''quantum foam''.

For some time the main approach to non-trivial spacetimes and Planck-scale effects has been Lorentz violation scenarios, which have been widely studied both theoretically and observationally. In this approach, Lorentz invariance is assumed to be broken at high energies, which introduces high-energy corrections to, for example, the dispersion relations of high-energy particles of cosmological origin. In recent years, the Rainbow Gravity framework~\cite{Magueijo:2002xx} has been given a lot of attention.~\cite{Ali:2014aba, Awad:2013nxa, Ling:2006az, Gwak:2016wmg, Khodadi:2016bcx, Ling:2006ba, Ling:2008sy, Khodadi:2015tsa, Liu:2015dza, Garattini:2013psa, Galan:2005ju, Galan:2004st, Ali:2014qra, Ashour:2016cay, Ali:2015iba}. This is a phenomenological approach based on Doubly Special Relativity (DSR), where the spacetime metric includes energy-dependent functions, and hence describes~\cite{Magueijo:2001cr, Magueijo:2002am} universes which evolve depending on the energy of the probe particle. With the correct choice of energy dependence, problems such as singularities may be avoided in Rainbow Gravity~\cite{Awad:2013nxa}. Exploring semiclassical or phenomenological theories of quantum gravity is of vital importance to understand the low-energy quantum gravitational regime and to reach an understanding of the underlying fundamental framework. 

It has been recently reported in~\cite{Lobo:2016lxm} that the Rainbow framework is suitable for exploring scenarios with broken Lorentz symmetry~\cite{Colladay:1998fq, Hees:2016lyw, Protheroe:2000hp, Ellis:2000sf, Stecker:2001vb, Stecker:2003pw, Coleman:1998ti, AmelinoCamelia:1997gz}. In the light of this, we present the following analysis which will be concentrated on the determination of the Lorentz violation energy scale for relativistic particles by the observational data from cosmology. 

This paper is organised as follows. In Section \ref{Basics} we briefly outline the formalism of Rainbow Gravity and Lorentz Invariance Violation (LIV) scenarios. In Section \ref{cosmology} we describe the modified homogeneous Friedmann universe in the Rainbow Gravity formalism. Section \ref{data} is dedicated to a statistical data analysis carried out which allows to constrain some rainbow parameters. In Section \ref{discussion} we interpret our results and present some concluding remarks. Unless explicitly stated, $c = \hbar = 1$, Greek indices $\mu, \nu = 0,1,2,3$, Roman indices $i, j, k = 1,2,3$, and the metric signature is $(-,+,+,+)$.

\section{Rainbow Gravity \& Doubly Special Relativity}
\label{Basics}

The key idea of Rainbow Gravity is the modification of the spacetime metric to include energy dependent functions $f_1(E)$ and $f_2(E)$~\cite{Magueijo:2002xx}, leading to a modified dispersion relation for relativistic particles of the form:
\begin{equation}
- E^2 f_1^2(E) + p^2 f_2^2(E) = m_0^2, 
\label{review:MDR}
\end{equation}
and position-space invariant of the form:
\begin{equation}
\text{ds}^2 = -\frac{(\text{dx}^0)^2}{f_1^2(E)} + \frac{(\text{dx}^i)^2}{f_2^2(E)}.
\label{review:metric}
\end{equation}
where $m_0$ is the rest mass of the particle. $x^0$ and $x^i$ are the time and space coordinates, respectively.
These functions are introduced by deforming the Lorentz group to include the Planck energy as a second invariant, using the formalism developed in Doubly Special Relativity (DSR) ~\cite{Magueijo:2002xx, Magueijo:2001cr, Magueijo:2002am}. By introducing the dilatation $D = p_\mu (\partial/\partial p_\mu)$, which preserves rotations but modifies boosts, the boost generators are deformed as follows:
\begin{equation}
K^i \equiv L_0^i + l_pp^iD \Rightarrow  K^i = U^{-1}L_0^{ i}U,
\label{review:generators}
\end{equation}
where $l_p$ is the Planck length and $L_0^{ i}$ are the conventional generators of the Lorentz group, $L_{\mu\nu} = p_\mu (\partial/\partial p^\nu) - p_\nu (\partial/\partial p^\mu)$~\cite{Magueijo:2001cr}. $U$ is a non-linear momentum map. $U$ in momentum space becomes:
\begin{equation}
U_\mu(E,p_i) = \left(U_0, U_i\right) = \left(Ef_1, p_if_2\right).
\label{review:U}
\end{equation}
By demanding plane-wave solutions to free field theories, $p_\mu x^\mu = p_0x^0 + p_ix^i$, the momentum map in position space is given by:
\begin{equation}
U^\alpha(x) = \left(U^0, U^i\right) = \left(\frac{t}{f_1}, \frac{x^i}{f_2}\right), 
\label{review:Upos}
\end{equation}
which leads to the position space invariant (and hence the metric):
\begin{multline}
s^2 = \eta_{\alpha\beta}U^\alpha(x) U^\beta(x) = -\frac{t^2}{f_1^2} + \frac{(x^i)^2}{f_2^2}
\Rightarrow g_{\alpha\beta}(E) = \text{diag}\left(-f_1^{-2}, f_2^{-2}, f_2^{-2}, f_2^{-2} \right),
\label{review:metric2}
\end{multline}
where $\eta_{\alpha\beta}$ are the components of the Minkowski metric.
In order to satisfy the correspondence principle, it is necessary to introduce a constraint on $f_1$ and $f_2$, namely
\begin{equation}
\lim_{E \to 0}f_k = 1, \, k = 1,2 \quad \Rightarrow \quad \lim_{E \to 0}g_{\mu\nu}(E) = \eta_{\mu\nu},
\label{review:constraint}
\end{equation}
which restores Minkowski space in the low-energy limit~\cite{Magueijo:2002xx}.
In DSR, invariants of the modified Lorentz group are accompanied by a singularity in the momentum map $U$~\cite{Magueijo:2001cr}. But in standard special relativity, the only energy invariant is the infinite one. Hence, to introduce a new invariant in the theory, the following relations must be fullfilled:
\begin{equation}
U(\tilde{E}) = \tilde{E}f_1(\tilde{E}) = \infty,
\label{using}
\end{equation}
where $\tilde{E}$ is some new invariant energy scale. This constraint, however, is not used by all authors; phenomenologically motivated rainbow functions $f_{1,2}$ which do not fulfill the criterium~\eqref{using} can be found in~\cite{Awad:2013nxa, AmelinoCamelia:1996pj} among others. 

The new metric $g_{\mu\nu}(E)$ defines a family of flat metrics parameterised by the energy $E$. Hence probe particles see ''different universes''; they measure different cosmological quantities and travel on different geodesics, but share the same set of inertial frames~\cite{Magueijo:2002xx}.

In order to apply DSR to cosmology it is necessary to find the Friedmann-Lemaitre-Robertson-Walker (FLRW) metric, as modified by Rainbow Gravity. Here the following system of units is implied: $dx^0 = c_0dt$, $c_0=1$, where $c_0$ is the low-energy limit of the energy-dependent speed of light, $c(E) \in [1,0]$. Now, we need to modify the FLRW metric. The resulting expression is:
\begin{equation}
\text{ds}^2 = -\frac{\text{dt}^2}{f_1^2(E)} + \frac{a^2(t)}{f_2^2(E)}\gamma_{ij}\text{dx}^i\text{dx}^j,
\label{flrw:metric}
\end{equation}
where $\gamma_{ij}$ represents the 3-metrics defined in Friedmann cosmology for the three different spacetime geometries ($K=0,\pm 1$), and $a(t)$ is the scale factor.
From the metric~\eqref{flrw:metric} we find the Einstein equations:
\begin{equation}
G_{\mu\nu}(E) = 8\pi G(E)T_{\mu\nu}(E) + g_{\mu\nu}(E)\Lambda(E),
\label{einstein}
\end{equation}
where all quantities now vary with energy. The tensorial quantities gain their energy dependence from the rainbow functions contained in the metric, whereas $G(E)$ and $\Lambda(E)$ get theirs from renormalisation group flow arguments, as outlined in~\cite{Magueijo:2002xx}. It is usually assumed that $G$ and $\Lambda$ have the same energy-dependence:
\begin{equation}
\begin{cases}
G(E) = h^2(E)G_0\\
\Lambda(E) = h^2(E)\Lambda_0
\label{hs}
\end{cases}
\end{equation}
where the index $0$ indicates the standard table value. The function $h(E)$, which we will now call the 'scaling function' is constructed in such a way that the standard constants $G_0$, $\Lambda_0$ are recovered in the limit $E \to 0$. Such form of the $h$-dependence for the gravitational and cosmological constants allows the constancy of the vacuum energy density $\rho_\Lambda = \Lambda_0/8\pi G_0$.
\section{Lorentz Invariance Violation in Rainbow Gravity}
\label{cosmology}

\subsection{Lorentz Invariance Violation}
Motivated by the notion of quantum foam coined by Wheeler~\cite{Wheeler:1957mu}, it has been suggested in theories of quantum gravity that Lorentz symmetry breaks down at high energies and short timescales~\cite{Colladay:1998fq, AmelinoCamelia:1996pj}. A common approach when studying these effects from a phenomenological point of view is to assume an effective modified dispersion relation, manifesting itself at high energies ~\cite{AmelinoCamelia:1997gz, AmelinoCamelia:2008qg, Coleman:1998ti}. In relation to that we consider a modified dispersion relation which for massless particles (whom we study from now on) takes the form:
\begin{equation}
p^2 = E^2 \, \rightarrow \, p^2 = E^2\left[ 1 + f(E)\right],
\label{LIV}
\end{equation}
A modified dispersion relation such as the one in Eq.~\eqref{LIV} would lead to highly energetic particles travelling slower or faster (depending on the quantum gravitational model) than their low-energy counterparts. For studies on Lorentz violation and possible observational tests, see~\cite{AmelinoCamelia:2008qg, Mattingly:2005re, Amelino-Camelia:2016wpo, Rosati:2015pga, Jacob:2008bw, Kifune:1999ex, Fairbairn:2014kda, Colladay:1998fq, Tasson:2016xib}.

In the framework of Lorentz Violation, it is often assumed that $f(E)$ in Eq.~\eqref{LIV} can be expressed in a series expansion at low energies ($E \ll E_{c}$)~\cite{AmelinoCamelia:1997gz, AmelinoCamelia:1996pj, 0004-637X-535-1-139}:
\begin{equation}
f(E) = \chi_1 \left(\frac{E}{E_{c}}\right)^1 + \chi_2 \left(\frac{E}{E_{c}}\right)^2 + \mathcal{O}\left[\left(\frac{E}{E_{c}}\right)^3 \right],
\end{equation}
where $E_{c}$ is the energy scale at which Lorentz violating effects become strong, and the couplings 
$\chi_n = \pm 1$ ($n=1,2$) are determined by the dynamical framework being studied. It is also assumed that the effects of Lorentz violation enter in either a linear or a quadratic term, and thus the low-energy approximation of $f(E)$ can be written as~\cite{AmelinoCamelia:1997gz}:
\begin{equation}
f(E) \approx \chi_n\left(\frac{E}{E_{c}}\right)^n .
\end{equation}
The modified dispersion relation in the present scenario then reads as:
\begin{equation}
\label{p2}
p^2 \approx E^2\left[1 + \chi_n \left(\frac{E}{E_{c}}\right)^n\right],
\end{equation}
which leads to a speed of light (or any other massless particle)~\cite{AmelinoCamelia:1997gz}:
\begin{equation}
c(E) = \frac{\partial E}{\partial p} \approx 1-\chi_n\left(\frac{E}{E_{c}}\right)^n, \quad E\ll E_{c},
\end{equation}
which changes its value as in VSL theories~\cite{AM99,VG,BM,KL15}. 

In quantum foam scenarios, the non-trivial features of spacetime at the Planck-scale are expected to slow particle propagation, and hence we will take $\chi_n = 1$ from now on.

It is now important to make the connection between this framework and Rainbow Gravity. In the latter, the invariant energy scale is the Planck scale. This is the energy scale which all observers agree on, and hence we identify $E_c = E_{Pl}$. Secondly, what we are ultimately interested in is the minimum energy which is needed for a massless particle to be subject to Lorentz violating effects. Hence, we will build a cosmological model in this framework and constrain the energy $E$ against data. Since no compelling evidence for Lorentz violation has yet been presented, the energy scale for Lorentz violation, $E_{LV}$, must be larger than the energy $E$. Hence, the only constraints we will be able to obtain will be lower limits.

\subsection{Simple Lorentz Invariance Violating Cosmological Framework}

It was recently reported in~\cite{Lobo:2016lxm} that the Rainbow formalism is suitable for describing Lorentz Violating scenarios~\cite{Colladay:1998fq, Hees:2016lyw, Protheroe:2000hp, Ellis:2000sf, Stecker:2001vb, Stecker:2003pw, Coleman:1998ti, AmelinoCamelia:1997gz}. It was shown that even though the Poisson bracket between the deformed boost and the flat-space limit Hamiltonian vanishes, $\{\mathcal{N}, \mathcal{H}\} = 0$, the Rainbow line element~\eqref{review:metric2} is not invariant under the same boost. The authors of~\cite{Lobo:2016lxm} remark that this makes vector norms non-invariant and makes it impossible to define local invariant observers, which makes it necessary to break Lorentz invariance~\cite{Lobo:2016lxm}. In the light of this, we present below a concatenation of Lorentz violation phenomenology and the Rainbow formalism, and we show that it is possible to combine the two in a consistent and logical way.

It is now possible to write down the Friedmann equation as follows~\cite{Magueijo:2002xx}:
\begin{equation}
\left(\frac{\dot{a}}{a}\right)^2 = \frac{8\pi G(E) \rho}{3f_1^2(E)} - \frac{K}{a^2}\left[\frac{f_2(E)}{f_1(E)}\right]^2 + \frac{\Lambda(E)}{3f_1^2(E)}, 
\label{friedmann}
\end{equation}
and the acceleration equation becomes:
\begin{equation}
\frac{\ddot{a}}{a} = - \frac{4\pi G(E)(3p + \rho)}{3f^2_1(E)} + \frac{\Lambda(E)}{3f_1^2(E)}.
\label{accel}
\end{equation}
Combining Eq.~\eqref{friedmann} and Eq.~\eqref{accel} yields the conservation equation, which is {\it independent} of the rainbow functions:
\begin{equation}
\dot{\rho} = 3\frac{\dot{a}}{a}(\rho + p).
\label{cons}
\end{equation}
The fact that the conservation equation does not include extra energy dependence from the rainbow functions is a clear advantage of this framework, since it implies that there is no dissipation of energy. Comparing the Lorentz Violation and Rainbow dispersion relations~\eqref{LIV} and~\eqref{review:MDR} and matching coefficients, it is possible to identify the following:
\begin{equation}
f_1(E) = \sqrt{1 + \left(\frac{E}{E_{Pl}}\right)^n}, \quad f_2(E) = 1
\end{equation}
From the dispersion relation~\eqref{LIV} and the correspondence principle it is possible to extract that $\lim_E \to 0, f(E) = 0$, which means that the map $U$ satisfies Eq.~\eqref{using}.

In order to calculate any useful cosmological quantities, it is neccessary to define $h(E)$. There are several suggestions in the literature, and the following two will be investigated. One suggestion comes from the field of \emph{varying constants cosmology}, where the running of physical constants is used to solve cosmological issues such as singularities.
In analogy with~\cite{Dabrowski:2012eb}, we suggest here that the evolution takes the following novel form:
\begin{equation}
G(E) = \left(1 - \frac{E}{E_{Pl}}\right)^{-1} G_0.
\label{h2}
\end{equation}
Comparing~\eqref{hs} and~\eqref{h2}, $h(E)$ is found to be (we will denote the first case $h_-$):
\begin{equation}
h_-(E) = \left(1 - \frac{E}{E_{Pl}}\right)^{-1/2}.
\label{myh}
\end{equation}

Another suggestion for the form of $h(E)$ can be found in~\cite{Khodadi:2015tsa}, and in analogy with this we suggest the following:
\begin{equation}
h_+(E) = \sqrt{1+ \left(\frac{E}{E_{Pl}}\right)^4}.
\label{hKhodadi}
\end{equation}

Choosing to look at a matter dominated universe with cosmological constant, $\rho = \rho_m$ and $\Lambda \neq 0$, the following solution to Eq.~\eqref{friedmann} is found:
\begin{equation}
a(t) = a_0 \left(\frac{\Omega_m}{\Omega_\Lambda}\right)^{1/3}\left[\sinh{\frac{3}{2}\sqrt{\Omega_\Lambda}\frac{h_\pm(E)}{f_1(E)}H_0t}\right]^{2/3},
\label{scalef1} 
\end{equation}
where $a_0$ is the present day value of the scale factor.
It is easy to see that~\eqref{scalef1} takes the standard form when $E \to 0$, so $h_\pm(E) \to 0$, which satisfies the correspondence principle.

As an example, we show here the case of $h_-(E) = \left(1-E/E_{Pl}\right)^{-1/2}$.
Using the rainbow function~\eqref{myh} in~\eqref{scalef1}, with $n = 2$, which in Lorentz violating scenarios is referred to as \emph{quadratic} Lorentz violation, the following result is obtained:
\begin{figure}[t]
\begin{center}
\includegraphics[width=.8\textwidth]{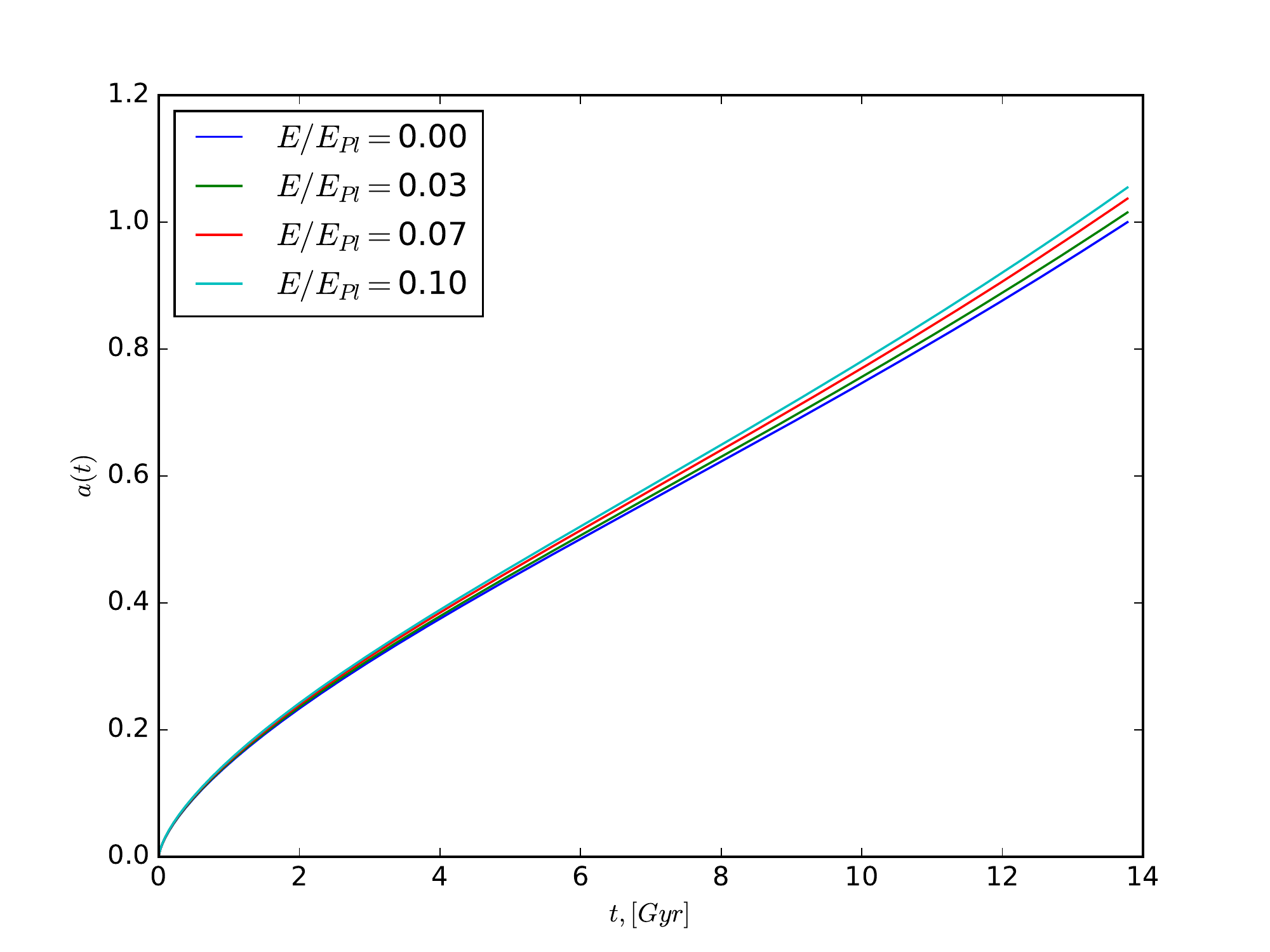}
\caption{The modified Friedmann scale factor for rainbow function~\ref{myh} with $n=2$ for probe particles of different energies.}
\label{scaleplot1}
\end{center}
\end{figure}
In Figure~\ref{scaleplot1}, the scale factors for the different probe energies clearly separate after $2-3$ Gyr, and the \emph{rainbow} in Rainbow Gravity can be clearly seen. Linear Lorentz violation, $n=1$ produces results which are difficult to distinguish when plotted. This is rather counterintuitive, as one would expect the less suppressed case ($n=1$) to be more important phenomenologically. The explanation to this lies in the function $h_-(E)$ which contains a minus sign in the denominator. Because of this the ratio $h_-/f_1$ contains terms such as $$1-\epsilon+\epsilon-\epsilon^2 = 1-\epsilon^2$$ in the case for $n=1$. (Here, $\epsilon = E/E_{Pl}$). The minus sign in $h_-$ causes this cancellation. For $n=2$, the corresponding term is $1-\epsilon$, when $\epsilon \ll 1$. Hence $h_-/f_1$, and more importantly, its derivative, will always be smaller for $n=1$ than $n=2$. This accounts for the somewhat surprising behavior of Figure 1.

In the more general case, when all the contributions to the energy density are taken into account, the Friedmann equation takes the following form:
\begin{equation}
\left(\frac{\dot{a}}{a}\right)^2 = \frac{8\pi G_0}{3} \frac{h_\pm^2(E)}{f_1^2(E)}\rho_c\left[\Omega_m \left(\frac{a_0}{a}\right)^3 + \Omega_{\text{rad}}\left(\frac{a_0}{a}\right)^4 + \Omega_\Lambda + \Omega_{k}\left(\frac{a_0}{a}\right)^2\frac{1}{h_\pm^2(E)}\right], 
\label{friedmann:realistic}
\end{equation}
where 
\begin{equation}
\Omega_{\{m, \text{rad}, \Lambda, k\}} = \frac{\rho}{\rho_c} = \frac{8\pi G}{3H_0^2}\rho_{\{m, \text{rad}, \Lambda, k\}},
\label{Omegas}
\end{equation}
are the energy density parameters (for matter, radiation, dark energy, and curvature) as measured today and $\rho_c$ is the conventional critical energy density, $\rho_c = 3H_0^2/8\pi G $. The extra factor on $\Omega_{k}$ comes from the definition of the curvature energy contribution:
\begin{equation}
\frac{h_{\pm}^2(E)}{f_1^2(E)}\frac{8\pi G}{3}\rho_{k} = -\frac{K}{a^2f_1^2(E)}
\label{omega:curv}
\end{equation}

\section{Constraints from Data}
\label{data}

\subsection{The method}
In this section, the expression $\dot{a}/a$ is denoted $H$.
In order to estimate the magnitude of the energy $E$ embedded in the rainbow functions $f_1(E), f_2(E)$, and $h_\pm(E)$, we used a large updated cosmological data set. The data used includes; expansion rates of elliptical and lenticular galaxies, Type Ia Supernovae, Baryon Acoustic Oscillations, Cosmic Microwave Background and priors on the Hubble parameter. For simplicity, all expressions below are expressed with zero spatial curvature ($\Omega_k = 0$). However, in the parameter estimation data analysis, $\Omega_k$ is left as a free parameter, and thus all equations extend to the more general case of~\cite{Hogg}. Hence, we use the following expression for the comoving distance:
\begin{equation}\label{eq:DM_k}
D_{M}(z) = \begin{cases} \frac{D_{H}}{\sqrt{\Omega_{k}}} \sinh \left( \sqrt{\Omega_{k}} \frac{D_{C}(z)}{D_{H}} \right) &\mbox{for } \Omega_{k} > 0\\
D_{C}(z) &\mbox{for } \Omega_{k} = 0 \\
\frac{D_{H}}{\sqrt{|\Omega_{k}|}} \sin \left( \sqrt{|\Omega_{k}|} \frac{D_{C}(z)}{D_{H}} \right) &\mbox{for } \Omega_{k} < 0 \, ,
\end{cases}
\end{equation}
where $D_{H} = c_{0} / H_{0}$ is the Hubble distance, $D_{C}(z) = D_{H} \int^{z}_{0} dz' / \mathcal{E}(z')$ is the line-of-sight comoving distance, and $\mathcal{E}(z) = H(z)/H_{0}$. $\Omega_k$ is the dimensionless curvature density parameter. Also, luminosity distance ($D_L(z)$) and angular diameter distance ($D_A(z)$) are given by:
\begin{eqnarray}
D_{L}(z) &=& (1+z) D_{M}(z)\; , \\
\label{angdist1}D_{A}(z) &=& \frac{D_{M}(z)}{1+z} \; .
\end{eqnarray}

\subsubsection{Hubble data}
For Hubble parameter data, we use the compilation from~\cite{Moresco15}, estimated from the evolution of elliptical and lenticular galaxies at redshifts $0<z<1.97$. The expression for $\chi^2_H$ in this case reads as:
\begin{equation}\label{eq:hubble_data}
\chi^2_{H}= \sum_{i=1}^{24} \frac{\left( H(z_{i},\boldsymbol{\theta})-H_{obs}(z_{i}) \right)^{2}}{\sigma^2_{H}(z_{i})} \; ,
\end{equation}
where $\boldsymbol{\theta}$ is a vector containing the cosmological parameters (including $E$), $H_{obs}(z_{i})$ are the measured values of the Hubble constant and $\sigma_{H}(z_{i})$ are the corresponding observational errors. We will also add a prior obtained from the Hubble constant in~\cite{Bennett14}, $H_0=69.6 \pm 0.7$ km s$^{-1}$ Mpc$^{-1}$.

\subsubsection{Type Ia Supernovae}
We used the JLA compilation (Joint Light-Curve Analysis) data for Type Ia supernovae (SneIa)~\cite{JLA} at redshifts $0<z<1.39$. In this case, the $\chi_{SN}^2$ is:
\begin{equation}
\chi^2_{SN} = \Delta \boldsymbol{\mu} \; \cdot \; \mathbf{C}^{-1}_{SN} \; \cdot \; \Delta  \boldsymbol{\mu} \; ,
\end{equation}
where $\Delta\boldsymbol{\mu} = \mu_{theo} - \mu_{obs}$ is the difference between theoretical and observational values of the distance modulus $\mu$. Here $\mathbf{C}_{SN}$ is the total covariance matrix. 
The distance modulus is defined as:
\begin{equation}\label{eq:m_jla}
\mu(z,\boldsymbol{\theta}) = 5 \log_{10} [ D_{L}(z, \boldsymbol{\theta}) ] - \alpha X_{1} + \beta \mathcal{C} + \mathcal{M}_{B} \; .
\end{equation}
Here, $X_1$ characterises the shape of the supernova light-curve, $\mathcal{C}$ is the colour, and $\mathcal{M}_B$ is a nuisance parameter~\cite{JLA}, which together with the weighting paramters $\alpha$ and $\beta$ are included in $\boldsymbol{\theta}$. $D_{L}$ is the luminosity distance, which is given by:
\begin{equation}\label{eq:dL}
D_{L}(z, \boldsymbol{\theta})  = \frac{1+z}{H_{0}}\int_{0}^{z} \frac{\mathrm{d}z'}{\mathcal{E}(z',\boldsymbol{\theta})} \; .
\end{equation}
Here, and only in the Supernova analysis, do we specify $H_0 = 70$ km/s Mpc$^{-1}$~\cite{JLA}.

\subsubsection{Baryon Acoustic Oscillations}
For Baryon Acoustic Oscillations (BAO), the total $\chi^2$ function is given by:
\begin{equation}
\chi^2_{BAO} = \Delta \boldsymbol{\mathcal{F}}^{BAO} \; \cdot \; \mathbf{C}^{-1}_{BAO} \; \cdot \; \Delta  \boldsymbol{\mathcal{F}}^{BAO} \; ,
\end{equation}
where $\boldsymbol{\mathcal{F}}^{BAO}$ differs from survey to survey. In this case, we used the WiggleZ Dark Energy Survey with redshifts $z=\{0.44,0.6,0.73\}$~\cite{WiggleZ}. 
For our purposes, the quantities to be considered are the acoustic parameter and the Alcock-Paczynski distortion parameter. The acoustic parameter is defined as follows:
\begin{equation}\label{eq:AWiggle}
A(z, \boldsymbol{\theta}) = 100  \sqrt{\Omega_{m} \, h^2} \frac{D_{V}(z,\boldsymbol{\theta})}{z} \, ,
\end{equation}
and the Alcock-Paczynski parameter reads as:
\begin{equation}\label{eq:FWiggle}
F(z, \boldsymbol{\theta}) = (1+z)  D_{A}(z,\boldsymbol{\theta})\, H(z,\boldsymbol{\theta}) \, ,
\end{equation}
where $D_{A}$ is the angular diameter distance, which is Eq.~(\ref{angdist1}) in the case of $\Omega_k = 0$:
\begin{equation}\label{eq:dA}
D_{A}(z, \boldsymbol{\theta} )  = \frac{1}{H_{0}} \frac{1}{1+z} \ \int_{0}^{z} \frac{\mathrm{d}z'}{\mathcal{E}(z',\boldsymbol{\theta})} \; ,
\end{equation}
and $D_{V}$ is the geometric mean of the physical angular diameter distance $D_A$ and the Hubble function $H(z)$. It reads as:
\begin{equation}\label{eq:dV}
D_{V}(z, \boldsymbol{\theta} )  = \left[ (1+z)^2 D^{2}_{A}(z,\boldsymbol{\theta}) \frac{z}{H(z,\boldsymbol{\theta})}\right]^{1/3}.
\end{equation}

Included in the Baryon Acoustic Oscillation analysis is also data from Sloan Digital Sky Survey (SDSS-III) Baryon Oscillation Spectroscopic Survey (BOSS) DR$12$~\cite{SDSS12}. It can be written as:
\begin{equation}
D_{M}(z) \frac{r^{mod}_{s}(z_{d})}{r_{s}(z_{d})} \qquad \mathrm{and} \qquad H(z) \frac{r_{s}(z_{d})}{r^{mod}_{s}(z_{d})} \,
\end{equation}
Here, $r_s(z_d)$ represents the sound horizon at the \emph{dragging redshift} $z_d$. $r^{mod}_s(z_d)$ is the same horizon, but evaluated for a given cosmological model. Here, it is used that $r^{mod}_s(z_d) = 147.78$ Mpc as in~\cite{SDSS12}. A good approximation of the sound horizon can be found in~\cite{Eisenstein}:
\begin{equation}\label{eq:zdrag}
z_{d} = \frac{1291 (\Omega_{m} \, h^2)^{0.251}}{1+0.659(\Omega_{m} \, h^2)^{0.828}} \left[ 1+ b_{1} (\Omega_{b} \, h^2)^{b2}\right]\; ,
\end{equation}
where
\begin{eqnarray}
b_{1} &=& 0.313 (\Omega_{m} \, h^2)^{-0.419} \left[ 1+0.607 (\Omega_{m} \, h^2)^{0.6748}\right], \nonumber \\
b_{2} &=& 0.238 (\Omega_{m} \, h^2)^{0.223}.
\end{eqnarray}
 
The sound horizon $r_s$ can then be defined as:
\begin{equation}\label{eq:soundhor}
r_{s}(z, \boldsymbol{\theta}) = \int^{\infty}_{z} \frac{c_{s}(z')}{H(z',\boldsymbol{\theta})} \mathrm{d}z'\, ,
\end{equation}
where the sound speed is given by:
\begin{equation}\label{eq:soundspeed}
c_{s}(z) = \frac{1}{\sqrt{3(1+\overline{R}_{b}\, (1+z)^{-1})}} \; ,
\end{equation}
and
\begin{equation}
\overline{R}_{b} = 31500 \Omega_{b} \, h^{2} \left( T_{CMB}/ 2.7 \right)^{-4}\; ,
\end{equation}
with $T_{CMB} = 2.726$ K.

To finish off the Baryon Acoustic Oscillation analysis, we also considered data from the Quasar-Lyman $\alpha$ Forest from Sloan Digital Sky Survey - Baryon Oscillation Spectroscopic Survey DR$11$~\cite{Lyman}:
\begin{eqnarray}
\frac{D_{A}(z=2.36)}{r_{s}(z_{d})} &=& 10.8 \pm 0.4\; , \\
\frac{1}{H(z=2.36) r_{s}(z_{d})}  &=& 9.0 \pm 0.3\; .
\end{eqnarray}

With these different contributions, the total $\chi^2$ for Baryon Acoustic Oscillations will be $\chi^{2}_{BAO} = \chi^{2}_{WiggleZ} + \chi^{2}_{BOSS} + \chi^{2}_{Lyman}$.

\subsubsection{Cosmic Microwave Background}
In this analysis, we write the $\chi^2$ for the Cosmic Microware Background (CMB) in the following way:
\begin{equation}
\chi^2_{CMB} = \Delta \boldsymbol{\mathcal{F}}^{CMB} \; \cdot \; \mathbf{C}^{-1}_{CMB} \; \cdot \; \Delta  \boldsymbol{\mathcal{F}}^{CMB} \; .
\end{equation}
Here, $\boldsymbol{\mathcal{F}}^{CMB}$ is a vector quantity given in~\cite{WangWang}, which summarises the information available in the full power spectrum of the Cosmic Microwave Background, as presented in the 2015 Planck data release~\cite{planck2015}. $\boldsymbol{\mathcal{F}}^{CMB}$ contains the Cosmic Microwave Background shift parameters and the baryonic density parameter. 
The shift parameters read as:
\begin{eqnarray}
R(\boldsymbol{\theta}) &\equiv& \sqrt{\Omega_m H^2_{0}} r(z_{\ast},\boldsymbol{\theta}) \nonumber \\
l_{a}(\boldsymbol{\theta}) &\equiv& \pi \frac{r(z_{\ast},\boldsymbol{\theta})}{r_{s}(z_{\ast},\boldsymbol{\theta})}\, ,
\end{eqnarray}
whereas the baryonic density parameter is simply $\Omega_b \, h^{2}$. As previously mentioned, $r_{s}$ is the comoving sound horizon at the photon-decoupling redshift $z_{\ast}$, which is given by~\cite{Hu}:
\begin{equation}{\label{eq:zdecoupl}}
z_{\ast} = 1048 \left[ 1 + 0.00124 (\Omega_{b} h^{2})^{-0.738}\right] \left(1+g_{1} (\Omega_{m} h^{2})^{g_{2}} \right) \, ,
\end{equation}
with:
\begin{eqnarray}
g_{1} &=& \frac{0.0783 (\Omega_{b} h^{2})^{-0.238}}{1+39.5(\Omega_{b} h^{2})^{-0.763}}\; , \\
g_{2} &=& \frac{0.560}{1+21.1(\Omega_{b} h^{2})^{1.81}} \, ;
\end{eqnarray}
and $r$ is the comoving distance:
\begin{equation}
r(z, \boldsymbol{\theta_{b}} )  = \frac{1}{H_{0}} \int_{0}^{z} \frac{\mathrm{d}z'}{\mathcal{E}(z',\boldsymbol{\theta})} \mathrm{d}z'\; .
\end{equation}

With all the abovementioned contributions to the total $\chi^2$, the function to minimise finally reads as: $\chi^2_{tot} = \chi^{2}_{H_{0}} + \chi^{2}_{H} + \chi^{2}_{SN} + \chi^{2}_{WiggleZ} + \chi^{2}_{BOSS} + \chi^{2}_{Lyman} + \chi^{2}_{CMB}$. Since the functions $f_1(E)$ and $h(E)$ will be expressed explicitly, the vector $\boldsymbol{\theta}$ will be written as $\boldsymbol{\theta} = \{\Omega_{m}, \Omega_{b}, \Omega_{k}, h, \alpha, \beta, E\}$.


We now want to find the set of parameters $\boldsymbol{\theta}$ that best fit the data set, we used a Markov-Chain Monte Carlo (MCMC) method, which was evaluated on the CI\'S computer cluster. The parameters are completely unconstrained but are given initial guesses, which speed up computation if they are chosen well. For visualisation, the Python package \texttt{corner} was used~\cite{corner}. During every step in the computation, the MCMC method calculates the $\chi^2$ mentioned above, and in the end returns the parameter set which minimised the $\chi^2$ function. This way, we are able to glean information about the posterior probability distribution without knowing it explicitly.

\subsection{Two specific choices of the scaling function $h_\pm(E)$}

The analysis described above was carried out for the two choices of the function $h_\pm(E)$ in Eq.~\eqref{myh} and Eq.~\eqref{hKhodadi} and limits on $E$ were derived for both linear and quadratic Lorentz violation ($n=1,2$). The results are stated in Table~\ref{tab:1}. In order to obtain these results, we employed an MCMC method, in which we ran three chains of $10^5$ steps each, to obtain bounds on the energy $E$.
These results are interpreted as follows; when constraining the energy $E$, we have looked for the values of $E$ which fit to our current understanding of the Universe, through the data available. Since Lorentz violating effects have not yet been observed, the energy scale $E_{LV}$ must lie {\it outside} of the likely range for $E$. As such, we obtain lower limits on $E_{LV}$ using the figures given in Table~\ref{tab:1}.
The limits placed correspond to the Grand Unified Theory (GUT) energy scale $E_{LV} \sim 10^{16}$ GeV at the $1\sigma$ limit and are even higher reaching $E_{LV} \sim 10^{17}$ GeV at $3\sigma$ limit which is very close to the so-called "Planck window".

In Table~\ref{tab:1} the case of $h_+(E)$ and $n=2$ is not included. Due to some artefact in the parametrisation this case contains {\it both} upper and lower limits on the energy scale $E$. We have discarded this case as it suggests we now live in a Lorentz violating era. Hence we deem it unphysical and do not consider it further.

At this stage it is very important to note that this is not the ''energy of spacetime'', but rather the energy scale of a probe particle travelling through spacetime and feeling a metric determined by its energy. This statement takes a central role in~\cite{Magueijo:2002xx}, where it is used to derive several modified cosmological quantities. In this paper, we interpret the limits obtained as decoupling limits, at which Lorentz violating effects become statistically significant. This is even clearer for the three models were we only obtained upper limits. This may be interpreted as a kind of arrival probability, and drops monotonically with energy. In analogy with the GZK cutoff, for example, we find this behaviour reasonable~\cite{gzk1,gzk2,gzk3}.

\renewcommand{\arraystretch}{2}
\begin{table}[t]
\begin{center}
\begin{tabular}{|c|c|}
\hline
\multirow{2}{*}{$h_-(E) = \sqrt{1-\frac{E}{E_{Pl}}}$} & $n=1:$ $0.0033$ $(1\sigma)$, $0.0076$ $(2\sigma)$, $0.0121$ $(3\sigma)$ \\ & $n=2:$ $0.0067$ $(1\sigma)$, $0.0152$ $(2\sigma)$, $0.0243$ $(3\sigma)$ \\
\hline
\multirow{1}{*}{$h_+(E) = \sqrt{1+\left(\frac{E}{E_{Pl}}\right)^4}$} & $n=1:$ $0.0068$ $(1\sigma)$, $0.0154$ $(2\sigma)$, $0.0262$ $(3\sigma)$\\
\hline
\end{tabular}
\caption{$1$, $2$, and $3\sigma$ constraints on the ratio $(E/E_{Pl})$ for linear and quadratic Lorentz violation ($n = 1, 2$) for the scaling function $h_-(E)$. For $h_+(E)$, only the case $n=1$ is included.}
\label{tab:1}
\end{center}
\end{table}

\subsection{Comparison with the $\Lambda$CDM model}
In our model, the decoupling energy scale from Lorentz violating effects leaves an imprint on the equations of cosmological evolution. As expected, this results in a different cosmological evolution compared to that of the $\Lambda$CDM. In order to quantify this difference, we notice that it is possible to write Eq.~\eqref{friedmann} in the following form:
\begin{equation}
\left(\frac{\dot{a}}{a}\right)^2 = \frac{8\pi G_0}{3}\rho_c\left[\Omega_m^\prime \left(\frac{a_0}{a}\right)^3 + \Omega_{rad}^\prime \left(\frac{a_0}{a}\right)^4 + \Omega_\Lambda^\prime + \Omega_k^\prime \left(\frac{a_0}{a}\right)^2\right],
\label{friedmann:rescaled}
\end{equation}
i.e. the standard form of the Friedmann equation. Here, the primed quantities are defined as (compare \eqref{friedmann:realistic} and \eqref{Omegas}):
\begin{equation}
\Omega_m^\prime = \frac{h_\pm^2(E)}{f_1^2(E)}\,\Omega_m, \quad \Omega_\Lambda^\prime = \frac{h_\pm^2(E)}{f_1^2(E)}\,\Omega_\Lambda, \quad \Omega_k^\prime = \frac{1}{f_1^2(E)}\,\Omega_k, \quad \Omega_{rad}^\prime = \frac{h_\pm^2(E)}{f_1^2(E)}\,\Omega_{rad} .
\label{rescaledOm}
\end{equation}
Besides, it is easy to notice from (\ref{rescaledOm}) that 
\begin{equation}
\frac{\Omega_m^{\prime}}{\Omega_m} = \frac{\Omega_{\Lambda}^{\prime}}{\Omega_{\Lambda}} = \frac{\Omega_{rad}^{\prime}}{\Omega_{rad}} ,
\end{equation}
and also that 
\begin{equation}
\Omega_{m} + \Omega_{rad} + \Omega_{k} + \Omega_{\Lambda} = \frac{h_\pm^2}{f_1^2}; \hspace{0.3cm} 
\Omega_{m}^{\prime} + \Omega_{rad}^{\prime} + \Omega_{k}^{\prime} + \Omega_{\Lambda}^{\prime} = 1.
\end{equation}

\begin{figure}[t]
\begin{center}
\includegraphics[width=.6\textwidth]{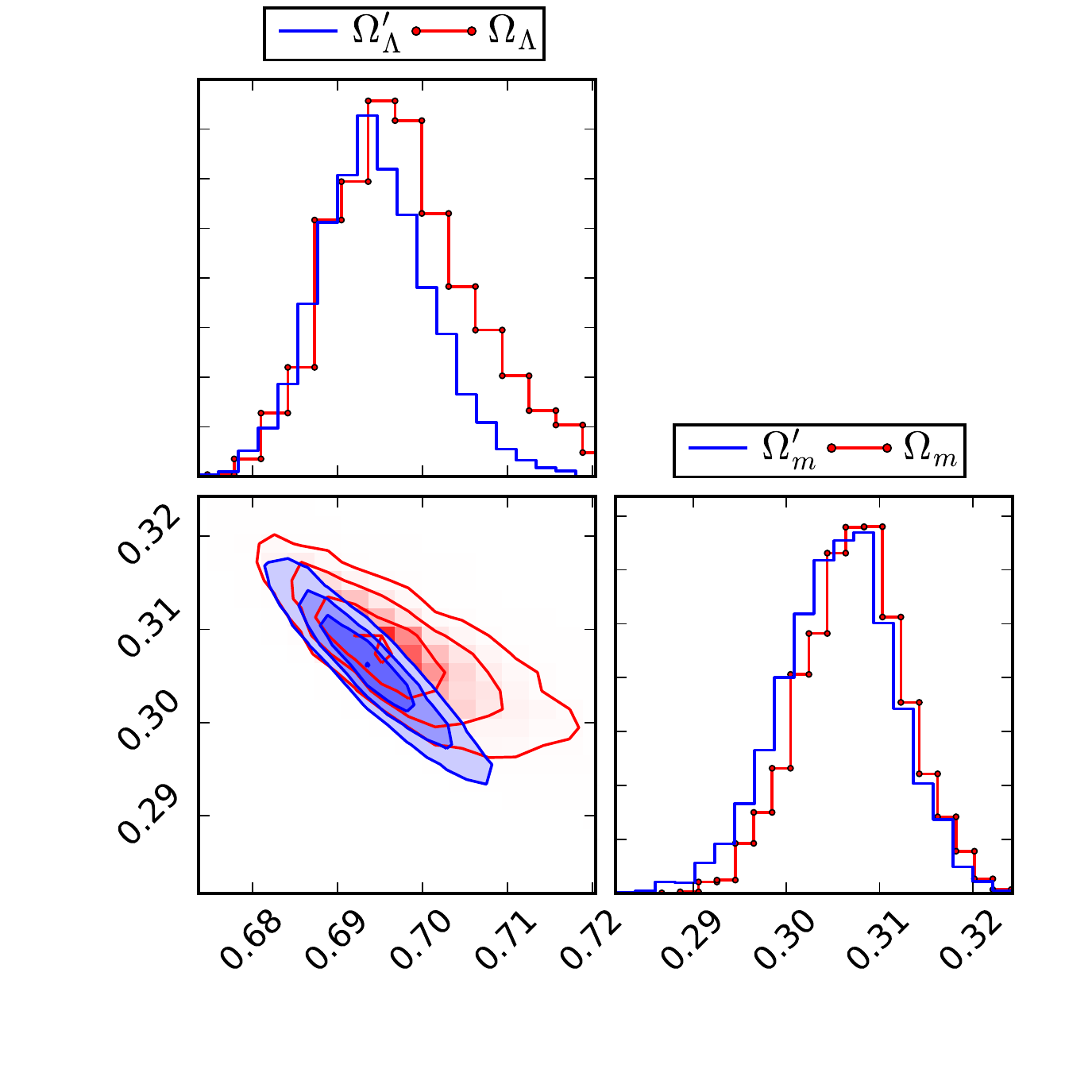}
\caption{One and two dimensional projections of the posterior probability distributions for linear Lorentz violation ($n=1$) and $h_-(E)=\sqrt{1-E/E_{Pl}}$. $\Omega_m$ and $\Omega_\Lambda$ (without primes) correspond to the energy densities in Eq.~\eqref{friedmann}, whereas $\Omega_m^\prime$ and $\Omega_\Lambda^\prime$ (with primes) are the rescaled quantities in Eq.~\eqref{friedmann:rescaled}. The histograms show the one dimensional marginalised distributions for the parameters independently, and the scatter plot shows the two dimensional parameter space.}
\label{E1n1_comp}
\end{center}
\end{figure}

As our analysis has provided bounds and estimates on the energy scale $E_{LV}$ as well as the energy densities $\Omega_X$, it is now a simple task to compare the primed and unprimed quantities. We present here the results for the model $h_-(E) = \sqrt{1-E/E_{Pl}}$ with $n=1$. In Figure~\ref{E1n1_comp} one sees the histograms for the matter and dark energy densities, both primed and unprimed.
From Figure~\ref{E1n1_comp} we can see that when rearranged to the standard Friedmann form, the primed quantities diminish in comparison to the unprimed ones. This was to be expected, as the ratio $h_\pm^2(E)/f_1^2(E)$ is consistently less than unity (in this model). As such, the imprint of the rainbow and scaling function on cosmological evolution can be thought of as mimicking dark energy in the sense that there is a weaker repulsion ($\Omega_{\Lambda}^{\prime} < \Omega_{\Lambda}$) accompanying weaker attraction ($\Omega_m^{\prime} < \Omega_m$) giving a net effect of a stronger global repulsion (acceleration). It is important to note that because of how the numerical analysis was carried out, the normalisation of primed and unprimed quantities are different.

\section{Discussion and Conclusions}
\label{discussion}
In this paper we have studied Lorentz symmetry violating scenarios which are predicted in the high-energy regime of some theories of quantum gravity. We have shown that it is possible to realise such scenarios within the framework of Rainbow Gravity due to modification of the dispersion relation by introducing new functions of particle energy $f_1(E)$ and $f_2(E)$. We have studied such a theory in the cosmological context assuming additionally the energy-dependence of the gravitational constant $G(E)$ and the cosmological constant $\Lambda(E)$ which change according to the scaling function $h_\pm(E)$. 

We have shown that it is possible to consistently express the low-energy limit of Lorentz violating theories within the framework of Gravity's Rainbow, when only one of the rainbow functions is non-trivial. We have proven that the Rainbow function $f_1(E)$ and the scaling functions $h_\pm(E)$ influence the evolution of the cosmological scale factor in the Friedmann equation. Our main point was to carry out a Markov-Chain Monte Carlo analysis in order to compare our theory with observational data such as: Hubble + Supernovae Type Ia + Baryon Acoustic Oscillations (Baryon Oscillation Spectroscopic Survey+Lyman $\alpha$) + Cosmic Microwave Background. Due to this we were able to constrain model parameters and in particular the energy scale $E_{LV}$ to be of the order of $10^{16}$ GeV at $1\sigma$ which is a Grand Unified Theory (GUT) energy scale up to $10^{17}$ GeV at $3\sigma$.

We suggest the interpretation of this energy as a Lorentz invariance decoupling scale since it is much higher than any observed particle energy. Just as the decoupling of the Cosmic Microwave Background in the early universe occurs at the recombination energy, the energy $E_{LV}$ may be interpreted as a decoupling energy from spacetime Lorentz violating effects. In the quantum foam picture, this occurs when the energy of a massless particle is too low to interact with the nontrivial spacetime, statistically. It may still happen through other mechanisms~\cite{Kifune:1999ex} and there are some possible observational signals of this (see for example~\cite{Fairbairn:2014kda}). 

We argue that the energy $E_{LV}$ should be viewed as the energy at which massless particles are decoupled from nontrivial background effects. This cutoff energy is generally assumed to be around the Planck energy, which this study indeed verifies. Moreover, the nontrivial structure of the quantum foam is expected to implicitly break Lorentz invariance, which can be modelled phenomenologically with a modified dispersion relation. This also fits well with our notion of $E_{LV}$, and as our assumptions on the structure of the function $f(E)$ stems from low-energy quantum gravity, our framework may be used for general quantum gravity phenomenology. It may be noted that our results are in agreement with some of the limits obtained in~\cite{HESS:2011aa}. It is also worth noting the behavior of Lorentz invariance hinted at in this paper is not a new idea; the notion of Lorentz symmetry being an \emph{emergent} symmetry is a key ingredient of Ho\v{r}ava Gravity~\cite{Horava:2008ih,Horava:2009if}, for example.

Several previous papers have investigated various aspects of the phenomenology of Rainbow Gravity (see for example~\cite{Hendi:2015vta,Bezerra:2017hrb,Torrome:2015cga,Gangopadhyay:2016rpl}). As a much expected consequence of quantum gravity the effects of Lorentz Violation should also be investigated.
Probing the behavior of symmetries at high energies is important in order to understand the limits of the current theories and to gain insight into what may lie beyond. Lorentz symmetry is one of those symmetries. However, as a fundamental ingredient of modern physics, it deserves thorough scrutiny.

\vspace{5mm}
{\bf Acknowledgements}\\
N.A.N. wishes to sincerely thank Vincenzo Salzano for very helpful discussions about parameter estimation and MCMC, and Viktor Svensson for contributing to important insights. M.P.D. wishes to thank Joao Magueijo for enlightening discussion about rainbow gravity. This work was financed by the Polish National Science Center Grant DEC-2012/06/A/ST2/00395. The use of the CI\'S computer cluster at the National Centre for Nuclear Research is gratefully acknowledged.






\bibliographystyle{elsarticle-num}
\bibliography{biblio.bib}
\end{document}